\documentclass[twocolumn]{aastex61}
\usepackage{natbib}
\usepackage{amsmath}

\newcommand{\logg}{\ensuremath{\log g}}

\newcommand{\feh}{\ensuremath{\protect\rm [Fe/H] }}
\newcommand{\teff}{T$_{\rm eff}$}
\usepackage{graphicx}
\usepackage{tablefootnote}
\usepackage{hyperref}
\usepackage[version=4]{mhchem}
\usepackage{txfonts}
\usepackage{comment}
\usepackage{tikz}
\usepackage{epsfig}
\usepackage[caption=false]{subfig}
\usepackage[flushleft]{threeparttable}
\usepackage{longtable}

\newcommand{\Ca}[1]{\ion{Ca}{#1}}

\newcommand\aastex{AAS\TeX}

\received{\today}
\shorttitle{\aastex\ Acurate Metallicities for very metal-poor stars from the Ca~II Infrared Triplet}
\shortauthors{Osorio et al.}

\begin{document}

 \title{Accurate Metallicities for Very Metal-Poor Stars from the Ca~II Infrared Triplet}

\author[0000-0001-5832-6933]{Yeisson Osorio}
\affil{Isaac Newton Group of Telescopes, Apto 321, E-38700 Santa Cruz de La Palma, Canary Islands, Spain \\}
\affiliation{Instituto de Astrof\'{\i}sica de Canarias,
              V\'{\i}a L\'actea, 38205 La Laguna, Tenerife, Spain\\}
\affiliation{Universidad de La Laguna, Departamento de Astrof\'{\i}sica, 
             38206 La Laguna, Tenerife, Spain \\}

\author[0000-0001-5200-3973]{David~S. Aguado}
\affiliation{Dipartimento di Fisica e Astronomia, Universitá degli Studi di Firenze, Via G. Sansone 1, I-50019 Sesto Fiorentino}
\affiliation{Institute of Astronomy, University of Cambridge, Madingley Road, Cambridge CB3 0HA, UK \\}

\author[0000-0002-0084-572X]{Carlos Allende Prieto}
\affil{Instituto de Astrof\'{\i}sica de Canarias,
              V\'{\i}a L\'actea, 38205 La Laguna, Tenerife, Spain\\}
\affiliation{Universidad de La Laguna, Departamento de Astrof\'{\i}sica, 
             38206 La Laguna, Tenerife, Spain \\} 

\author[0000-0001-8816-236X]{Ivan Hubeny}
\affil{Steward Observatory, University of Arizona, Tucson, USA}

\author[0000-0002-0264-7356]{Jonay~I. Gonz\'alez Hern\'andez}
\affil{Instituto de Astrof\'{\i}sica de Canarias,
              V\'{\i}a L\'actea, 38205 La Laguna, Tenerife, Spain\\}
 \affiliation{Universidad de La Laguna, Departamento de Astrof\'{\i}sica, 
             38206 La Laguna, Tenerife, Spain \\}
\email{yo@ing.iac.es}

\begin{abstract}
 The Ca~II H \& K lines are among the few features available to infer the metallicity of extremely metal-poor stars from medium-resolution spectroscopy. Unfortunately,  these lines can overlap with absorption produced in the  intervening interstellar medium,  introducing systematic errors in the derived metallicities. The strength of the Ca~II infrared triplet lines can also be measured at extremely low metallicities, and it is not affected by interstellar absorption, but suffers significant departures from Local Thermodynamical Equilibrium (LTE). We investigate the feasibility of adopting the Ca~II infrared triplet as a metallicity indicator in extremely metal-poor stars  using state-of-the art Non-LTE models including the most recent atomic data. We find that the triplet lines exhibit Non-LTE abundance corrections that can exceed 0.5\,dex. When interstellar absorption affecting the Ca~II resonance lines is accounted for  using  high-resolution observations, the agreement between Non-LTE abundances for the triplet and those for the resonance lines, with only minor departures from LTE, is excellent. Non-LTE effects strengthen the Ca~II IR triplet lines, facilitating measurements at very low metallicities, compared with LTE estimates, down to $\rm [Fe/H]=-6.0$. This result has important implications for the discovery of primitive stars in our Galaxy and others, since instruments are most sensitive at red/near-infrared  wavelengths, and tens of millions of spectra covering the Ca~II IR triplet will soon become available from the Gaia, DESI, WEAVE, and PFS missions.

\end{abstract}

\keywords{line: formation --- stars: population II ---  stars: abundances ---  stars: Atmospheres }

\section{Introduction} \label{sec:intro}
The first stars born in the Galaxy inherited the composition of the pristine gas in the early interstellar medium, with basically no metals. We do not yet know whether there is a metallicity threshold for the formation of low-mass stars \citep[e.g.][and references therein]{bro03, chi17, stacy16, magg18}, but there are low-mass stars  with metal-to-hydrogen ratios as low as $10^{-5}$ \citep[see, e.g.,][]{ishi14, komiya16}, and even lower ratios when calcium or iron abundances are taken as a proxy for metallicity. 

Very-old low-mass stars that have survived until today offer us a way to track the early chemical evolution of the Milky Way, and sample the nucleosynthetic yields of the first massive stars and their supernovae  \citep[e.g.,][]{umeda05, tilman18}. But to do so we need first to find suitable samples-- they are truly rare, with only half a dozen known at [Fe/H]$<-5$ \citep{chris04, fre05, kel14, boni15, agu18I, agu18II, nord19}, and second, learn to read their spectra, which challenge simplified models of stellar atmospheres.

At low metallicity the overall atmospheric structure is affected by the lack of metal-line absorption. This has been shown to have dramatic consequences for the formation of molecules \citep{GP2006}. Lower metal line opacity, together with a lower density due to a lower mean molecular weight. leads to a larger number of scatterings experienced by photons in spectral lines and consequently to more pronounced deviations from LTE.

At extremely low metallicities, stellar spectra, especially for warm F-type halo turnoff stars, appear void of metal lines, and only the abundances of a few elements can be measured. Searches for extremely metal-poor stars exploit the formidable strength of the Ca~II H \& K lines at 3934 and 3969~\AA, which can be detected down to  very low abundances at modest spectral resolution \citep[e.g.,][]{bee85,caffau13, agu16}. However, calcium atom in the warm interstellar medium, with typical temperatures in the range 6000--10000 K, are mainly sitting in the ground level of Ca~II, and {\it contribute absorption in the vicinity of the stellar} lines. The interstellar medium absorption may have multiple components from individual clouds, which are narrower and  usually shifted in velocity from the stellar lines. These contributions are hard to resolve and take into account when deriving stellar calcium abundances  from low- and medium-resolution spectra.

The Ca~II triplet lines at 8498, 8542 and 8662~\AA\ are somewhat weaker than the resonance lines, but still fairly strong and not affected by interstellar absorption. They are in a spectral window that is nearly optimal for CCD detector efficiency, and remain visible in a wide variety of spectral types, which has motivated their choice for the Gaia's Radial Velocity Spectrograph \citep{gaia_rvs}. Their use for metallicity estimation in extremely metal-poor stars has been hampered by the fact that the levels connected by the triplet are affected by departures from Local Thermodynamical Equilibrium (LTE), and a reliable description of the effect of inelastic collisions with atomic hydrogen was lacking and is critical. This issue has been addressed in the recent calculations by \citet{2018ApJ...867...87B}.

\cite{2019MNRAS.485.3527S} have studied in detail the formation of calcium lines in very metal-poor stars, upgrading the model from \cite{2007A&A...461..261M} including the recent data for collisions with hydrogen atoms for Ca~II from \cite{2018ApJ...867...87B} as well as those from \cite{2017ApJ...851...59B} for Ca~I. The main Non-LTE effect is an overpopulation of the lower level and photon losses in the cores of the triplet lines, which strengthens the lines. \cite{2019MNRAS.485.3527S} find that Non-LTE abundance corrections for the triplet depend mainly on line strength. 

We have updated our model atoms, including the best possible collisional and radiative data, and put then to the test using existing and new, unpublished, observations of several extremely metal-poor stars. Section \ref{obs} reports on the observational material. Section \ref{cal} gives an account of the atomic data, model atmospheres, and line formation calculations. The last two sections describe our results (\S \ref{res}) and conclusions (\S \ref{con}).

\section{Observations}
\label{obs}

There are 3 sets of observations used in this work: medium-resolution spectroscopy with ISIS, on the 4-m William Herschel Telescope (WHT), and high-resolution data obtained with HORuS, installed at the 10.4-m Gran Telescopio Canarias (GTC),  and UVES, in operation at the Very Large Telescope (VLT). \

The observations taken with ISIS span the ranges  3500--5200 \AA\ and 7420--9180 \AA and therefore cover both, the Ca~II H\&K lines and the IR triplet at low resolution. Unfortunately, the Ca~II H line (3968~\AA) is not resolved in the ISIS observations. HORuS covers approximately the range between 3800 and 6900 \AA, missing the infrared triplet, but allowing for a more detailed study of the H\&K lines since both lines are resolved at the spectral resolution of this instrument (about 25,000). The resolving power for the UVES observations is higher (R$\sim$45,000), and the spectral coverage is wider and includes the infrared triplet lines. More details are given below.

\subsection{HORuS} 

G64-12 was observed with HORuS \citep{2020MNRAS.498.4222T, 2021NatAs...5..105A} at GTC\footnote{http://research.iac.es/proyecto/abundancias/horus/index.php} on March 11 and July 2, 2019, HD122563 on January 13th 2020; and SDSS 134338 was observed on December 3, 2019. Data reduction was carried out with the {\tt chain}\footnote{Available from github.com/callendeprieto/chain}, which performs bias and flatfield correction, order tracing, extraction, and wavelength calibration. The observed spectra of the same target were combined in order to increase the signal-to-noise ratio (S/N). These spectra include some Ca~I lines as well.

\subsection{UVES}
The high resolution Ultraviolet and Visual Echelle Spectrograph \citep[UVES]{dek00} spectra of HE 1327$-$2326 and HD 122563 were extracted from the ESO archive\footnote{Based on data from the ESO Science Archive Facility. Program ID: 077.D-0035(A)}. Radial velocities for individual spectra were determined by cross correlation with a synthetic template, and after correcting for it, the exposures were co-added to obtain a final spectrum with a S/N of $\sim120$.  

\subsection{ISIS}
Both G64$-$12 and HE 1327$-$2326 were previously studied by \citet{agu17I}, using medium-resolution data taken  at  the  4.2\,m  William  Herschel Telescope (WHT) with the Intermediate dispersion Spectrograph and Imaging System spectrograph \citep[ISIS,][]{isiswht}. The setup included the R600B and R600R gratings with the default dichroic, allowing us to measure the Ca~II H\&K resonance doublet  and the Ca II infrared triplet. Each target was observed for a total of 3600\,s, leading to a S/N of about 300 and 130 at 3940 \AA\ and 8550 \AA, respectively.

\section{Calculations}
\label{cal}

 We carried out NLTE calculations with the latest version of {\tt Tlusty} (version~208\footnote{\hyperlink{link to iac cloud}{link to iac cloud}}), which can calculate NLTE populations in the trace-element approach using precomputed opacity tables, and {\tt Synspec} (version~54\footnote{Same link.}) which generates opacity tables and detailed LTE and NLTE synthetic spectra \citep{tlusty1,tlusty2,tlusty3,tlusty4}. 

The stellar parameters adopted for G64-12 and HE1327-2326 were taken from \cite{2006ApJ...639..897A}. In the case of SDSS134338, the parameters were calculated by some of us \citep{agu17I}. The parameters for HD 122563 were extracted from \cite{2015A&A...582A..49H}, where they obtained the parameters via bolometric flux and isochrone calibrations  \citep[see also, ][]{2016MNRAS.463.1518A}. Effects of the stellar parameters on our NLTE results are presented below.

All the synthetic spectra were calculated in the same fashion: Model atmospheres were computed afresh using the ATLAS9 \citep{2005MSAIS...8...14K} posix port from \cite{2007IAUS..239...71S}. For each model atmosphere an opacity table was constructed with {\tt Synspec}, adopting the same abundances as the model atmosphere, but removing the contribution from Ca and Mg. The metallicity indicator we used in the calculations is the usual, [Fe/H]. It is known that in metal poor stars [Ca/Fe]$\approx$0.4~dex but at [Fe/H] $\lesssim$ -4.0~dex, the scatter of [Ca/Fe] increase \citep[see e.g., ][]{2020A&A...642A..25F}. The (T, $\rho$) pairs in the opacity table match those of the atmosphere, avoiding interpolation. The tables include 100\,000 frequency points, sampled logarithmically in wavelength between 900 and 100\,000~\AA, and were used in {\tt Tlusty} for the calculation of the NLTE populations for Ca and Mg. Those populations were fed back to {\tt Synspec} to produce detailed NLTE and LTE spectra to be compared with observations\footnote{The model atmospheres could have been computed by Tlusty as well, using an analogous full opacity table, but due to historical reasons [and for a comparison to previous studies] we have opted to stick with the ATLAS code.}.

\addtolength{\tabcolsep}{-1pt}  
\begin{table*}
\caption{Stellar parameters for the synthetic atmospheric models used in our comparison with observations.}\label{tab:stars}
\centering
\begin{tabular}{l l c c c c c c c}\hline\hline
   &  & T$_{\rm eff}$ & $\log g$ & [Fe/H] & [C/Fe] & [O/Fe] & $v_{turb}$ & Parameters source \\
Target  & Instrument & (K) & (cm/s$^2$)  & (dex) & (dex) & (dex) & km/s & \\\hline
G64-12   & ISIS/HORuS   &  6390 &   4.4 &  -3.2  &  1.0 & 1.0  &  1.6 & \citet{2006ApJ...639..897A} \\ 
SDSS134338 & HORuS  &  5888 &   4.1 &  -3.3  &  1.0 & 1.0  &  1.5 & \citet{agu17I} \\
HE1327-2326 & ISIS/UVES &  6180 &   3.7 &  -5.6  &  4.0 & 3.0  &  1.7 & \citet{2006ApJ...639..897A} \\
HD 122563    & UVES &  4587 &   1.6 &  -2.6  &  1.0 & 1.0  & 1.0 & \citet{2016MNRAS.463.1518A} \\\hline
\end{tabular}
\end{table*}
\addtolength{\tabcolsep}{1pt}  

\subsection{Model atom}
 
We adopted the model atom described in \cite{CaPaperI}, with updates in the hydrogen interactions with Ca~I and II presented by  \cite{2018ApJ...867...87B}. This update includes charge exchange between protons and calcium atoms and between H and singly-ionized calcium. 

The model atom has 66 levels for Ca~I, 24 for Ca~II, and the ground level of Ca~III. These figures are different to those presented in \cite{CaPaperI} since, for this run, we merged all fine-structure components into single levels. This saves computing time without degrading accuracy\footnote{The calculations in \cite{CaPaperI} and other works showed that NLTE populations of fine-structure components of the same level share nearly identical departures from LTE, and can be treated as a single level.}.

The radiative data used were the same as in \cite{CaPaperI}. The transitions  involving fine-structure components were merged accordingly \citep[see][\S 2.1]{CaPaperI}. This leads to 912 and 89 bound-bound transitions for Ca~I and Ca~II, respectively.  

The electron collisional data are the same as in \cite{CaPaperI} i.e., \cite{2000asqu.book.....C,1970ARA&A...8..329B,Vrinceanu:2005em} for electron collisional ionisation, while for electron collisional excitation we adopted \cite{PhysRevA.99.012706,1962amp..conf..375S} for Ca~I and \cite{2007A&A...469.1203M} and \cite{1977MNRAS.179..275B} for Ca~II. 

In \cite{CaPaperI}, Hydrogen collisions were considered only for Ca~I and neglected for Ca~II. For Ca~I + H, they used \cite{2016PhRvA..93d2705B,2017PhRvA..95f9906B} and \cite{Kaulakys:1986tl}. In this work we also implemented charge exchange between Ca~I and excited H($n=2$); 
hydrogen collisional excitation of Ca~II (for transitions involving levels up to 4f~$^2$F$^*$) and charge exchange with Ca~II
\[
\text{\Ca{2}} + \text{H}  \Longleftrightarrow \text{\Ca{3}} + \text{H}^-.
\]
These new data come from \cite{2018ApJ...867...87B}. The importance for this work of these new processes lies on the fact that we focus our study on Ca~II features, and the new collisional data directly affect the statistical balance of the levels involved in these transitions. 

We also studied Mg and Ca inter-species NLTE effects by simultaneously calculating NLTE populations of Mg and Ca (neutral and singly-ionized), as in \cite{2020A&A...637A..80O}.

\section{Results}
\label{res}

For the three most metal-poor stars studied in this work, assuming LTE leads to overestimated calcium abundances. NLTE effects strengthen the Ca~II lines, making their detection easier at lower calcium abundances. We also found negligible inter-species NLTE effects between calcium and magnesium. Only for the coolest model (SDSS 134338, $T_{\rm eff}=$5888 K), the population of the highest levels of Ca~II and the ground level of Ca~III are perturbed by a few percent due to NLTE effects of Mg in the outermost atmospheric layers. Our observations do not include lines of Ca~II involving those levels, but inter-species effects with other ions should be studied.

Using the HORuS observations of G64-12 we derived the calcium abundance in this star from three transitions: the Ca~II H\&K lines  and Ca~I $\lambda$4227 (see Table \ref{tab:abun}). The first thing to notice is the consistency between the Ca~I and Ca~II NLTE abundances. In the LTE case, the abundances from the Ca~II resonance doublet are~$\approx$~0.1.~dex greater than the values from the Ca~I $\lambda$4227 transition.

We studied the effects of spectral resolution on the derived LTE abundances and NLTE abundance corrections for G64-12 and SDSS 134338. At the resolving power of HORuS ($R=25,000$), the  best-fitting NLTE profile for the Ca~II K line reproduces more closely the observations than the LTE one. This is no longer the case at the lower resolution of the ISIS data, which are fit equally well in LTE and NLTE. The NLTE correction for this line doubles between HORuS and ISIS resolution: while the NLTE abundances derived from fitting the line profile are the same, the inferred LTE abundance increases between the HORuS and the ISIS best fittings. This is simply the result of lower resolution spreading the NLTE effects affecting the line core over a wider range of frequencies, enhancing the systematic error in the LTE-based abundance.

\section{Formation depth and line formation, the case of HD122563.}

We included HD122563 in the analysis, the most metal rich star in our sample, as illustrated in table~\ref{tab:stars}. We found that the core of Ca-II H\&K,  Ca-I 4227~\AA\ and the Ca-II IR triplet lines form at layers beyond the limits of our model atmospheres ($log(\tau_{ross})=-6$) and special care must be taken for the derived abundances from these lines.  

The core of the H\&K and the lines form at the top layers of our model atmospheres, but we expect those regions to be less realistic due to NLTE effects on the structure of the atmospheres and temperature inversions which are not accounted for in standard models of late type stellar atmospheres. 

\begin{figure}
    \centering
    \begin{tikzpicture}
\node[anchor=south east, inner sep=0] (image) at (0,0) { \subfloat{\includegraphics[width=0.46\textwidth]{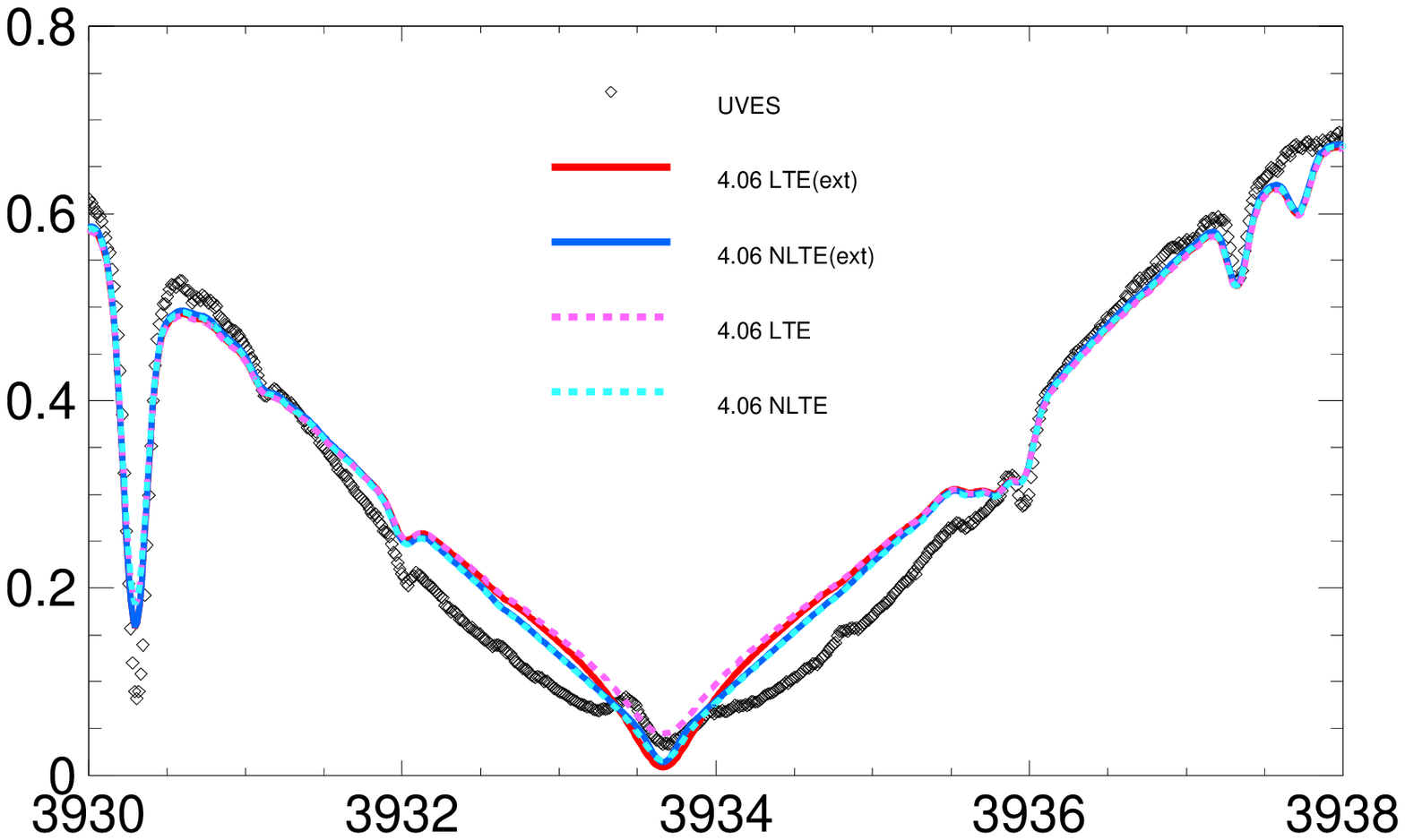}}};
    \draw [fill=white,white] (-0.242\textwidth,0.25\textwidth) rectangle (-0.15\textwidth,0.123\textwidth);
     \node[right] at (-0.24\textwidth,0.245\textwidth){HD122563};
     \node[right] at (-0.24\textwidth,0.220\textwidth)[scale=1.00]{4.06 LTE (ext)};
     \node[right] at (-0.24\textwidth,0.195\textwidth)[scale=1.00]{4.06 NLTE (ext)};
     \node[right] at (-0.24\textwidth,0.170\textwidth)[scale=1.00]{4.06 LTE};
     \node[right] at (-0.24\textwidth,0.145\textwidth)[scale=1.00]{4.06 NLTE};    \end{tikzpicture}
    \\
    \begin{tikzpicture}
\node[anchor=south east, inner sep=0] (image) at (0.47\textwidth,0) {  \subfloat{\includegraphics[width=0.47\textwidth]{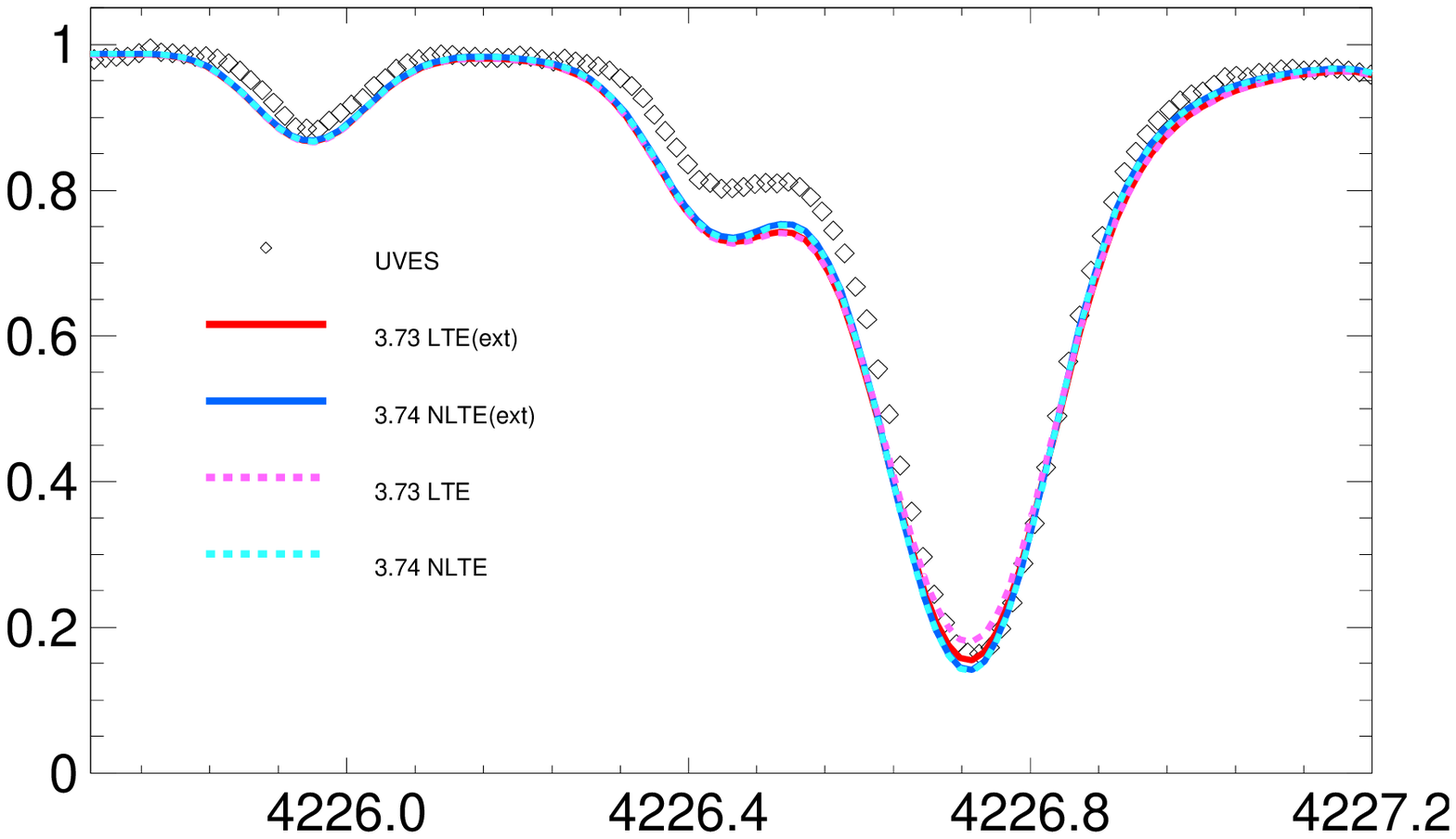}}};    
    \draw [fill=white,white] (0.2\textwidth,0.2\textwidth) rectangle (0.108\textwidth,0.08\textwidth);
     \node[right] at (0.11\textwidth,0.190\textwidth){HD122563};
     \node[right] at (0.11\textwidth,0.165\textwidth)[scale=1.00]{3.73 LTE (ext)};
     \node[right] at (0.11\textwidth,0.140\textwidth)[scale=1.00]{3.74 NLTE (ext)};
     \node[right] at (0.11\textwidth,0.115\textwidth)[scale=1.00]{3.73 LTE};
     \node[right] at (0.11\textwidth,0.090\textwidth)[scale=1.00]{3.74 NLTE};   
    \end{tikzpicture}
    \\
    \begin{tikzpicture}
    \node[anchor=south east, inner sep=0] (image) at (0.47\textwidth,0) { \subfloat{\includegraphics[width=0.47\textwidth]{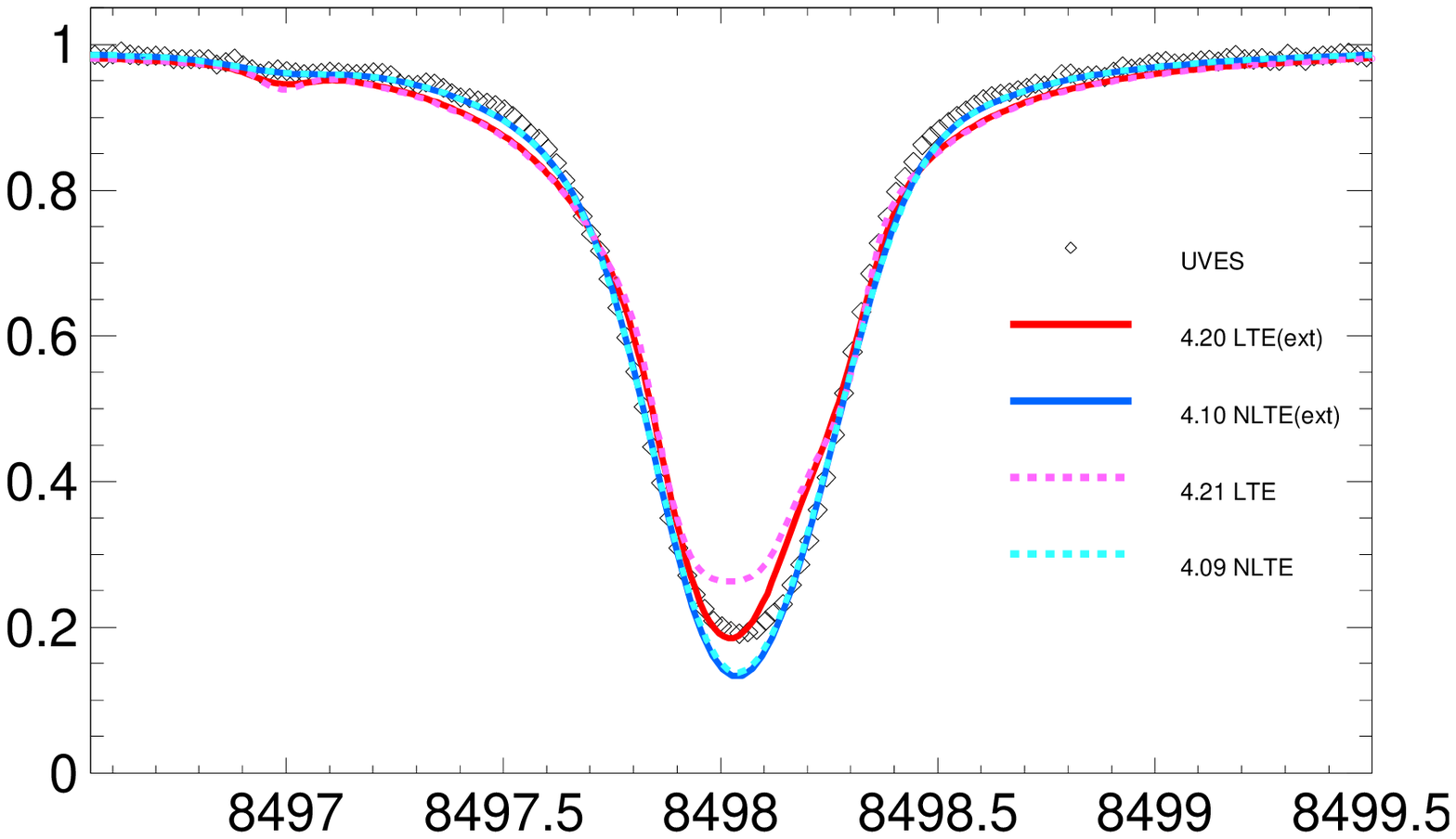}}};    
    \draw [fill=white,white] (0.435\textwidth,0.2\textwidth) rectangle (0.37\textwidth,0.08\textwidth);
     \node[right] at (0.355\textwidth,0.190\textwidth){HD122563};
     \node[right] at (0.365\textwidth,0.167\textwidth)[scale=1.00]{4.20};
     \node[right] at (0.365\textwidth,0.142\textwidth)[scale=1.00]{4.10};
     \node[right] at (0.365\textwidth,0.117\textwidth)[scale=1.00]{4.21};
     \node[right] at (0.365\textwidth,0.092\textwidth)[scale=1.00]{4.09};   
    \end{tikzpicture}
    \caption{Comparison between UVES observations of HD122563 and the best fit LTE (red) and NLTE (blue) spectra for the Ca II-K line (top), the Ca I 4226 (middle) and the IR triplet member Ca II 8498~\AA\ line (bottom). Solid lines use an extended model atmosphere (min $\log\tau_{ross}=-8.0$), light dashed lines goes up to the standard depth (min $\log\tau_{ross}=-5.0$).}
    \label{fig:HD122HK}
\end{figure}

The dependency of the calculated line profile on the extension of the model atmosphere used in the calculation is most clear in the Ca I 4227~\AA\ line, shown in the middle panel of fig. \ref{fig:HD122HK}. The core of this line is formed at $\log\tau_{ross}\sim-6.0$ in LTE and at $\log\tau_{ross}\sim-4.5$ in NLTE. The gf value of this particular line comes from \cite{Hulpke1964} with an error of around 3\% (0.013~dex for $\log{gf}$) which can not explain the difference between the synthetic profile and observations. 

NLTE corrections of strong lines computed by comparison of equivalent widths are sensitive to the extent of the stellar atmosphere because the core of the LTE profile of strong lines gets deeper with more extended atmospheres, which does not happen with the NLTE profile.

It is known that the LTE approximation gradually loses validity with atmospheric height until it is no longer useful in the highest layers. 
Lines with weak wings and with cores that form at very high atmospheric layers (like the Ca I 4227~\AA\ line) should not be used for abundance determination because the region where the core forms is not modelled correctly due to the extension of the model atmosphere and problems in the modelling of the highest layers of cool stellar atmospheres. These problems appear in both, 1D and 3D atmospheric models.

\section{Most extreme metal-poor stars}

For the other stars in this study, the cores of the Ca II H\&K, the Ca I 4227 and the Ca II IR triplet lines do not form at the highest layers and therefore the differences between the LTE and NLTE line profiles can not be attributed to the extent of the model atmospheres.

\def\vert{1}
\if\vert1
\begin{figure}[ht]       
    \hspace{-0.005\textwidth}
    \begin{tikzpicture}
    \node[anchor=south east, inner sep=0] (image) at (0,0) {
    \subfloat{\includegraphics[width=0.47\textwidth]{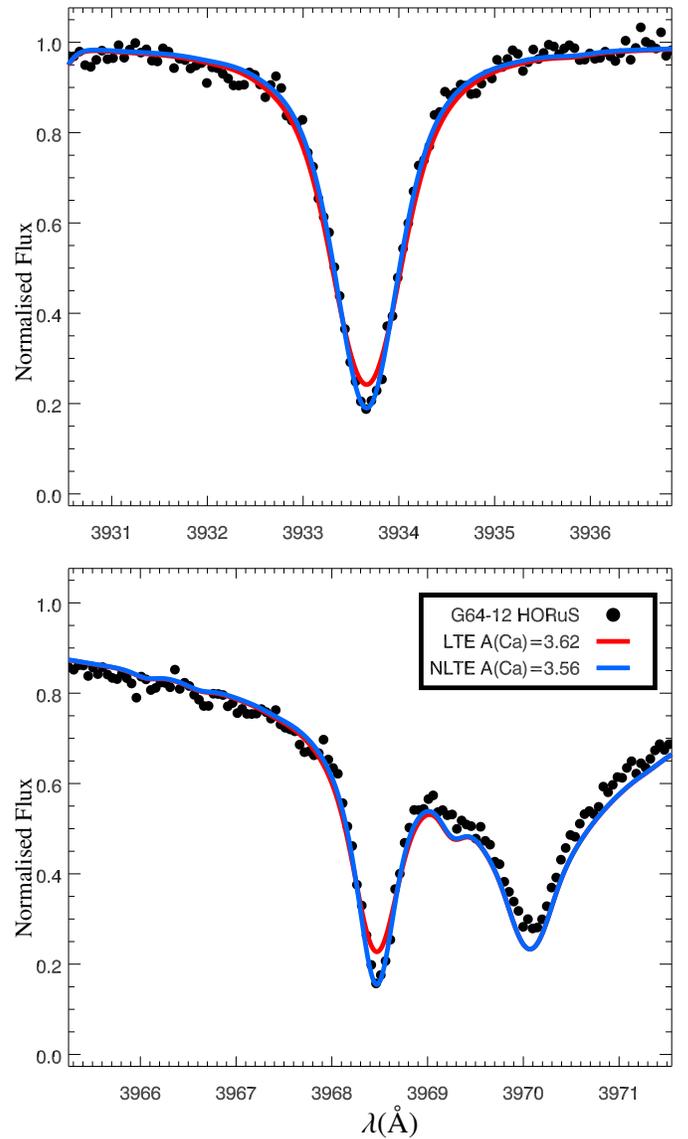}}};
    \node[rotate=90] at (-0.48\textwidth,0.18\textwidth)[scale=1.00]{Normalised Flux};
    \node[rotate=90] at (-0.48\textwidth,0.59\textwidth)[scale=1.00]{Normalised Flux};
    \node at (-0.21\textwidth,-0.015\textwidth)[scale=1.25]{$\lambda$(\AA)}; 
    \end{tikzpicture}
    \caption{Comparison of the \Ca{2} 3934~\AA\ and \Ca{2} 3969~\AA\ (H\&K) lines between HORuS observations of G64-12 (dots) and the best LTE(red) and NLTE(blue) fit. The Ca\ abundance used for each calculation is in the legend.}
    \label{fig:g6412_HK}
\end{figure}
\else 
\begin{figure*}[t]       
    \hspace{0.005\textwidth}
    \begin{tikzpicture}
    \node[anchor=south east, inner sep=0] (image) at (0,0) {
    \subfloat{\includegraphics[width=0.94\textwidth]{bbG61-12_HK_HORuS_horizontal.eps}}};
    \node[rotate=90] at (-0.97\textwidth,0.175\textwidth)[scale=1.25]{Normalised Flux};
    \node at (-0.21\textwidth,-0.015\textwidth)[scale=1.25]{$\lambda$(\AA)}; 
    \node at (-0.70\textwidth,-0.015\textwidth)[scale=1.25]{$\lambda$(\AA)}; 
    \end{tikzpicture}
    \caption{Comparison of the \Ca{2} 3934~\AA\ and \Ca{2} 3969~\AA\ (H\&K) lines between HORuS observations (dots) and the best LTE(red) and NLTE(blue) fit. The Ca\ abundance used for each calculation is in the legend.}
    \label{fig:g6412_HK}
\end{figure*}
\fi

\begin{table*}
    \centering
    \caption{Calcium abundances inferred for our target stars from the individual transitions}
    \label{tab:abun}
    \footnotesize
    \begin{tabular}{llllll l l l l l l l}\hline\hline
    \phantom{Starss}line(\AA) $\rightarrow$  & \multicolumn{2}{c}{\Ca{2}\,3934} & \multicolumn{2}{c}{\Ca{2}\,3968} & \multicolumn{2}{c}{\Ca{1}\,4227}  & \multicolumn{2}{c}{\Ca{2}\,8497}  & \multicolumn{2}{c}{\Ca{2}\,8542}  & \multicolumn{2}{c}{\Ca{2}\,8662}  \\
 Star        &   LTE    &  NLTE         &  LTE     &   NLTE        &  LTE      &    NLTE       &   LTE     &  NLTE         &   LTE     &  NLTE         &   LTE     &  NLTE       \\\hline
 HD122563 & 4.06(05) & 4.06(04) & & & 3.73(05) & 3.74(05) &    4.20(05) & 4.10(04) \\
 G64-12      & 3.64(06) & 3.60(04)      & 3.66(09) & 3.57(06)      & 3.53(12)  &   3.63(09)    &  4.08(12) & 3.63(09)      & 4.04(10)  & 3.61(11)      & 4.11(10)  & 3.64(12)      \\
 SDSS 134338  & 3.42(04) & 3.41(03)      &          &               & 3.35(07)  &   3.40(07)    &  3.82(08) & 3.38(06)      & 3.80(06)  & 3.41(06)      & 3.85(11)  & 3.42(11)      \\
 HE1327-2326 & 1.31(11) & 1.15(09)      & 1.32(30) & 1.11(21)      &           &               &           &               & 1.22(24)  & 1.16(19)      &          &             \\\hline

    \end{tabular}
   
\end{table*}

\subsection{Ca~II H \& K lines}

The Ca~II resonance lines are resolved in the HORuS observations of G64-12. The line profiles are nicely reproduced in NLTE, while in LTE the models miss absorption in the line cores, as illustrated in Figure \ref{fig:g6412_HK}.

The Ca~I $\lambda$ 4266 line has been reported to give discrepant abundances compared to other Ca~I lines when the line is strong (with an equivalenth width $W_{\lambda}>$50 m\AA) \citep{2007A&A...461..261M,2017A&A...605A..53M,2019MNRAS.485.3527S}. This disagreement was reported both in LTE and NLTE analyses.  We do not find this issue in our analysis of G64-12, where $W_{\lambda} \simeq$~100~m\AA. We derive a NLTE abundance correction of 0.1~dex, and our NLTE determination is consistent with the values from the other calcium lines, as can seen in Table \ref{tab:abun}. We believe the reported discrepancy is related to the limited extension of the atmospheric models used by \citet{2007A&A...461..261M,2017A&A...605A..53M} and \citet{2019MNRAS.485.3527S}. The core of the resonance lines forms in very high layers of the photosphere, and the Kurucz model atmospheres we adopt typically reach optical depths two orders of magnitude smaller than the MARCS model atmospheres used in those previous studies, as discussed by \cite{CaPaperI}.

At the resolving power of HORuS, we cannot find signs of contamination from the interstellar medium (ISM) in the spectra of G64-12. For this star the Ca~II resonance lines and the IR triplet treated in NLTE lead to consistent calcium abundances, while an LTE analysis of the triplet results in a discrepancy of 0.4~dex.

\begin{figure*}[t]
    \begin{tikzpicture}
    \node[anchor=south east, inner sep=0] (image) at (0,0){
    \subfloat{\includegraphics[height=0.35\textwidth]{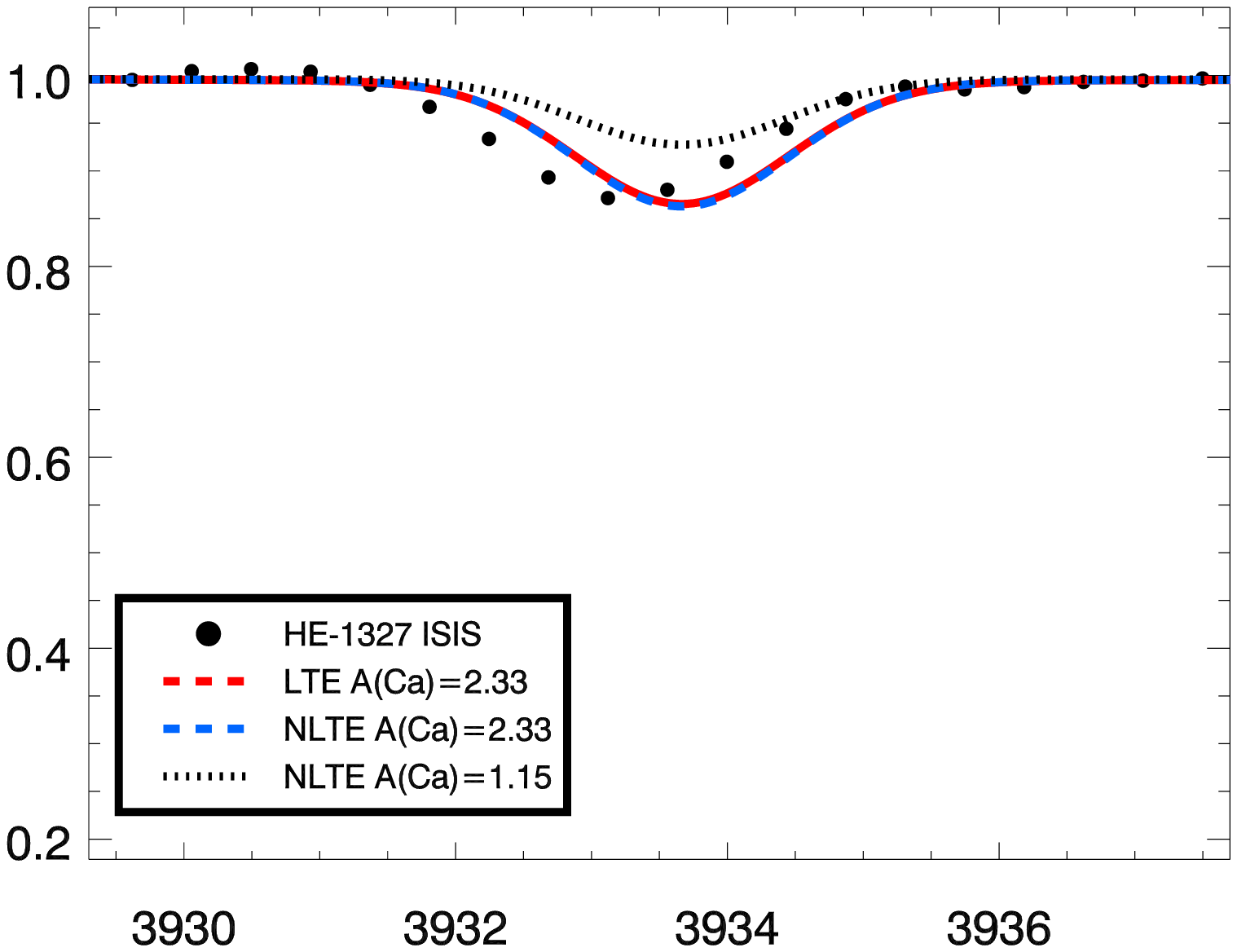}}
    };  

    \node at (-0.22\textwidth,-0.01\textwidth)[scale=1.3]{$\lambda$(\AA)};
    \node[rotate=90] at (-0.49\textwidth,0.17\textwidth)[scale=1.3]{Normalized flux};
    \end{tikzpicture} 
    \hspace{0.005\textwidth}
        \begin{tikzpicture}
    \node[anchor=south east, inner sep=0] (image) at (0,0){
    \subfloat{\includegraphics[height=0.35\textwidth]{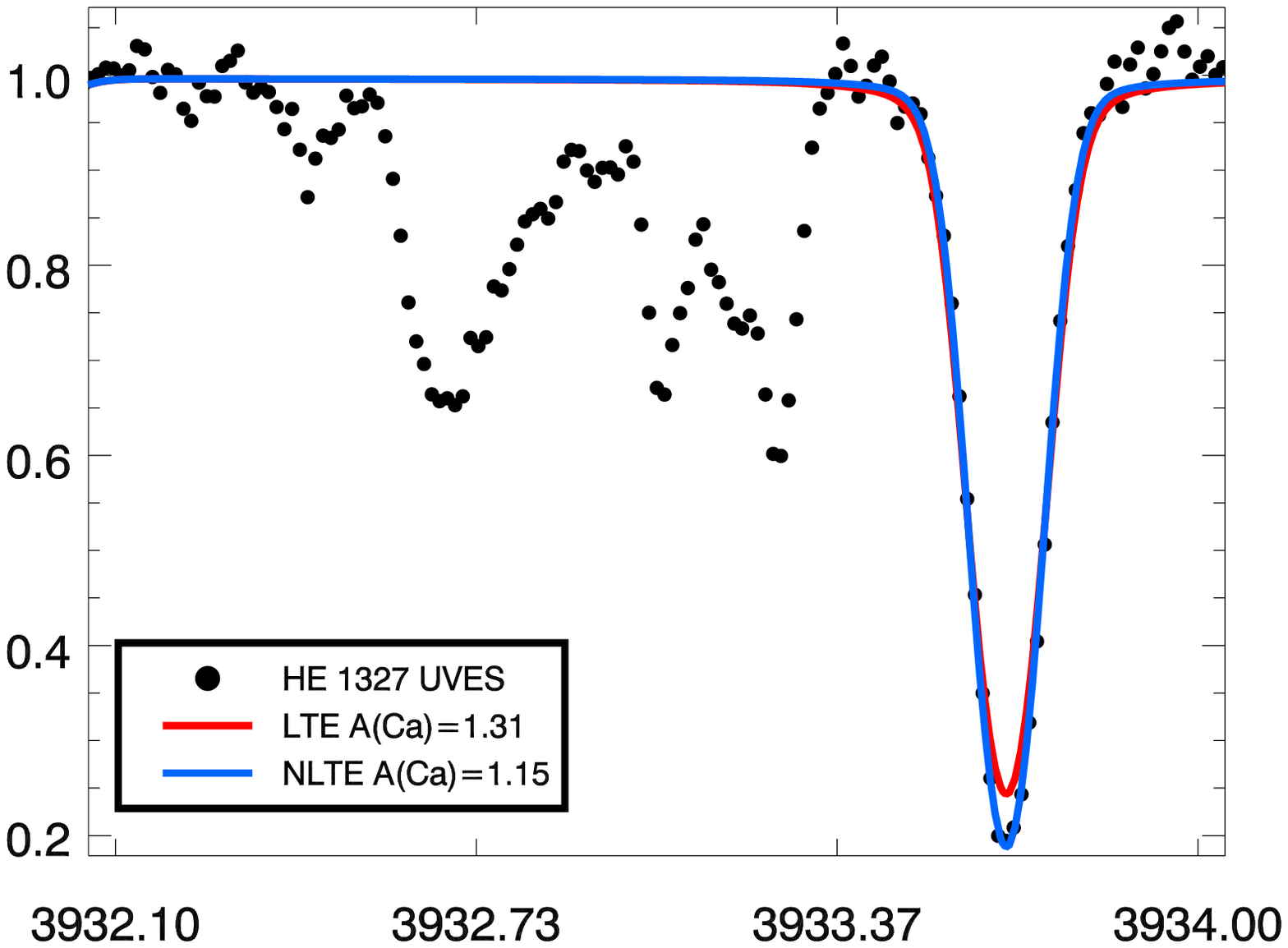}}
    };
    \node at (-0.23\textwidth,-0.01\textwidth)[scale=1.3]{$\lambda$(\AA)};
    \end{tikzpicture}

    \caption{Ca LTE and NLTE synthetic profile vs observations with ISIS (R=2\,000, left) and UVES (R=40\,000, right) of the Ca~II K line in HE~1327-2326. The inner plot in the left figure is a zoom on the Ca~II 3930~\AA\ line and the ISIS observations.The Ca abundance used in the synthetic spectra calculations are written in the legends.}
    \label{fig:he1327b}
\end{figure*}

The UVES observations of HE~1327 show at least three lines blue-shifted from the stellar Ca~II H and K lines, as illustrated for the K line in the right-hand panel of Fig. \ref{fig:he1327b}. The LTE and NLTE best fittings for these lines give an NLTE abundance correction\footnote{defined here as $\Delta$A(Ca) = A(Ca)$_{\rm{NLTE}}-$A(Ca)$_{\rm{LTE}}$.} of $-0.15$ dex for this star. At the lower resolution of the ISIS data, the interstellar components are blended with the stellar K line (left-hand panel). Ignoring the ISM contribution would lead to a higher calcium abundance by more than 1.0~dex. 

The dotted line in the left-hand panel of Fig. \ref{fig:he1327b} shows the NLTE line profile of the K line using the NLTE abundance derived from the UVES observations (right panel) at the resolution of ISIS. By looking at the UVES observations, one would expect the $W_{\lambda}$ of the resulting blend at ISIS resolution to have around twice the value of the stellar component alone. If the K line were in the linear region of the curve of growth in this star we expect an increase in abundance of $\sim0.5$~dex, The strength of the line at the resolution of UVES shows that this line is saturated. 

\subsection{Ca~II IR triplet}

The IR triplet lines do not suffer from ISM contamination, but the levels involved in the IR triplet transitions (3d~$^2$D - 4p~$^2$P$^o$), the first two excited levels of Ca~II,  are more suitable to exhibit populations that differ from LTE predictions. There are ISIS  observations of these lines for all three stars studied in this work. The LTE and NLTE values are significantly different, and in all cases our derived NLTE abundances exhibit less dispersion than the LTE ones. 

The NLTE corrections for the  IR triplet in G64-12 and SDSS 134338 are $\sim-$0.4~dex. The NLTE abundances are consistent with those from other calcium lines. In contrast, the derived LTE abundances for these lines are higher.  

There are multiple factors that complicate the abundance determination from the Ca~II IR-Triplet lines. Two of those lines ($\lambda\lambda$8497, 8662) are on the blue wings of hydrogen lines that are stronger than the calcium lines. The  limited signal-to-noise ratio of the ISIS data makes the same two lines lie within the noise level. Thus, for HE~1327, we derived the calcium abundance only from the 8542~\AA\ line. The NLTE abundance correction in this case is small, of about 0.05~dex. 

We performed NLTE calculations on an small grid ($4500 \leq $\teff$ \leq 6000, \Delta=500$ K ; $1.0 \leq \logg \leq 4.0, \Delta=1.0$~dex ; $-6.0 < \feh < -2.5, \Delta=0.5$~dex) and present together with this work abundance corrections $\Delta A(Ca)$ for the triplet IR lines (see Table \ref{tab:acorr}).

\section{Conclusions}
\label{con}

Absorption produced in the interstellar medium (ISM) can lead to systematic errors in the determination of stellar abundances of calcium from the Ca~II resonance lines. The Ca~II IR triplet offers an alternative way of measuring the metallicity of these stars, free from ISM absorption, but afflicted by significant departures from LTE. Earlier analysis reported inconsistencies between the abundances derived from these transitions, as well as those from neutral calcium lines, and we consider whether recent updates in the data available for the relevant physical processes have improved the situation.

We perform Mg+Ca multi-species NLTE (NLTE-m) calculations, including the most relevant collisional processes. The new H+Ca$^{+}$  collisional data from \cite{2018ApJ...867...87B} do have significant effects on our NLTE calculations. NLTE effects strengthen  the core of the Ca~II infrared triplet lines, reducing the inferred abundance, and getting it into agreement with the values from the resonance lines, mostly free from NLTE effects, whenever the latter can be isolated from ISM contributions. 

We conclude that our state-of-the-art NLTE calculations, similarly to those recently published by \citet{2019MNRAS.485.3527S}, are mature enough to standardize the use of the Ca~II IR triplet to infer reliably the calcium abundance in very metal-poor stars from medium-resolution observations. Furthermore, we make public all the necessary data and tools for anyone to repeat and extend our calculations to any other star.

Together with this paper, we release scripts, line lists, model atoms,  and the versions of {\tt Tlusty/Synspec} deployed for this work \citep{tlusty4}. Additional calculations are possible by running these scripts.

This paper is focused on NLTE, however, it is well-known that 3D effects can be important for the accurate inference of atmospheric parameters and the abundances of some species at low metallicity 
\citep{2005ARA&A..43..481A, 2013MNRAS.429..126R, 2016MNRAS.463.1518A, 2019MNRAS.488L.109N}. G64-12 in particular was analysed by \citep{2012MNRAS.427...27B} in NLTE using MARCS and {\it horizontaly averaged} 3D atmospheres, and this star and HD 122563 were studied by \citep{2016MNRAS.463.1518A} et al. using full 3D NLTE. \citet{2013MNRAS.429.3645S} include a useful discussion of the influence of LTE and NLTE model atmospheres in the Ca near-IR triplet lines at low metallicity.

Our conclusions, together with the actual data and tools to extend our calculations to other stars, have important implications for the discovery of extremely metal-poor stars in the new future. As demonstrated in this letter, the triplet can be reliably used to infer stellar metallicities in these stars when departures from LTE are accounted for. In addition, \citet{2017A&A...597A...6N} have shown that 3D effects are not likely to change the  calcium abundances derived in 1D from Ca~II lines in extremely metal-poor stars by more than 0.1 dex.

Several massive spectroscopic surveys of the Milky Way covering the Ca~II infrared lines are  underway or close to start (SDSS, LAMOST, DESI, WEAVE, PFS,  ...). Next year the 3rd Gaia Data Relase will make public millions of $R\simeq 11,500$ spectra covering the Ca~II IR triplet. The tools presented in this paper can be use to mine with confidence these data for identifying new extremely metal-poor stars.

\begin{acknowledgements}
    We thank the referee for his/her valuables remarks and suggestions during the referring process. 
    Y.O., C.A.P. and J.I.G.H. acknowledge financial support from the Spanish Ministry of Science and Innovation (MICINN) under the FEDER Agreement INSIDE-OOCC (ICTS-2019-03-IAC-12) and from the Spanish MICINN project AYA2017-86389-P.
     DA thanks the Leverhulme Trust for financial support.
     DA acknowledges support from ERC Starting Grant NEFERTITI H2020/808240.
    Based on observations made with the William Herschel Telescope (WHT) and the Gran Telescopio Canarias (GTC), installed in the Spanish Observatorio del Roque de los Muchachos of the Instituto de Astrofísica de Canarias, in the island of La Palma.
\end{acknowledgements}

\begin{longtable}{l l l c c c}
\caption{NLTE abundance corrections of Ca derived from the three lines of the Ca~II IR triplet for an small grid. The first three columns are \teff, \logg\ and \feh respectively; the last three columns are the abundance correction $\Delta$A(Ca) for each of the IR Ca~II triplet lines.} \label{tab:acorr} \\

Teff & logg & Fe/H & 8498\AA & 8542\AA & 8662\AA \\\hline
\endfirsthead
Teff & logg & Fe/H & 8498\AA & 8542\AA & 8662\AA \\\hline
\endhead

\hline
\endfoot

\hline\hline
\endlastfoot

4500 & 1.0 & -2.5 & -0.171 & -0.224 & -0.175 \\
4500 & 1.0 & -3.0 & -0.158 & -0.122 & -0.135 \\
4500 & 1.0 & -3.5 & -0.152 & -0.114 & -0.140 \\
4500 & 1.0 & -4.0 & -0.222 & -0.191 & -0.213 \\
4500 & 1.0 & -4.5 & -0.276 & -0.298 & -0.311 \\
4500 & 1.0 & -5.0 & -0.349 & -0.430 & -0.424 \\
4500 & 1.0 & -5.5 & -0.412 & -0.554 & -0.526 \\
4500 & 1.0 & -6.0 & -0.383 & -0.609 & -0.575 \\
4500 & 2.0 & -2.5 & -0.195 & -0.290 & -0.222 \\
4500 & 2.0 & -3.0 & -0.155 & -0.143 & -0.146 \\
4500 & 2.0 & -3.5 & -0.112 & -0.083 & -0.100 \\
4500 & 2.0 & -4.0 & -0.202 & -0.159 & -0.183 \\
4500 & 2.0 & -4.5 & -0.246 & -0.247 & -0.262 \\
4500 & 2.0 & -5.0 & -0.291 & -0.353 & -0.349 \\
4500 & 2.0 & -5.5 & -0.295 & -0.454 & -0.436 \\
4500 & 2.0 & -6.0 & -0.283 & -0.499 & -0.454 \\
4500 & 3.0 & -2.5 & -0.208 & -0.423 & -0.293 \\
4500 & 3.0 & -3.0 & -0.161 & -0.201 & -0.180 \\
4500 & 3.0 & -3.5 & -0.160 & -0.148 & -0.157 \\
4500 & 3.0 & -4.0 & -0.188 & -0.168 & -0.180 \\
4500 & 3.0 & -4.5 & -0.218 & -0.226 & -0.233 \\
4500 & 3.0 & -5.0 & -0.223 & -0.297 & -0.293 \\
4500 & 3.0 & -5.5 & -0.220 & -0.354 & -0.327 \\
4500 & 3.0 & -6.0 & -0.222 & -0.323 & -0.278 \\
4500 & 4.0 & -2.5 & -0.232 & -0.546 & -0.371 \\
4500 & 4.0 & -3.0 & -0.178 & -0.281 & -0.234 \\
5000 & 1.0 & -2.5 & -0.272 & -0.248 & -0.244 \\
5000 & 1.0 & -3.0 & -0.280 & -0.214 & -0.253 \\
5000 & 1.0 & -3.5 & -0.283 & -0.255 & -0.292 \\
5000 & 1.0 & -4.0 & -0.293 & -0.337 & -0.351 \\
5000 & 1.0 & -4.5 & -0.287 & -0.427 & -0.418 \\
5000 & 1.0 & -5.0 & -0.348 & -0.560 & -0.522 \\
5000 & 1.0 & -5.5 & -0.336 & -0.624 & -0.571 \\
5000 & 1.0 & -6.0 & -0.276 & -0.610 & -0.537 \\
5000 & 2.0 & -2.5 & -0.310 & -0.294 & -0.294 \\
5000 & 2.0 & -3.0 & -0.318 & -0.261 & -0.303 \\
5000 & 2.0 & -3.5 & -0.329 & -0.296 & -0.331 \\
5000 & 2.0 & -4.0 & -0.317 & -0.364 & -0.382 \\
5000 & 2.0 & -4.5 & -0.302 & -0.458 & -0.444 \\
5000 & 2.0 & -5.0 & -0.271 & -0.523 & -0.479 \\
5000 & 2.0 & -5.5 & -0.220 & -0.544 & -0.481 \\
5000 & 2.0 & -6.0 & -0.194 & -0.456 & -0.352 \\
5000 & 3.0 & -2.5 & -0.283 & -0.344 & -0.304 \\
5000 & 3.0 & -3.0 & -0.288 & -0.259 & -0.279 \\
5000 & 3.0 & -3.5 & -0.297 & -0.260 & -0.287 \\
5000 & 3.0 & -4.0 & -0.280 & -0.303 & -0.318 \\
5000 & 3.0 & -4.5 & -0.227 & -0.361 & -0.349 \\
5000 & 3.0 & -5.0 & -0.154 & -0.395 & -0.357 \\
5000 & 3.0 & -5.5 & -0.136 & -0.366 & -0.285 \\
5000 & 3.0 & -6.0 & -0.131 & -0.233 & -0.179 \\
5000 & 4.0 & -2.5 & -0.240 & -0.422 & -0.314 \\
5000 & 4.0 & -3.0 & -0.228 & -0.252 & -0.243 \\
5000 & 4.0 & -3.5 & -0.223 & -0.220 & -0.228 \\
5000 & 4.0 & -4.0 & -0.047 & -0.080 & -0.070 \\
5000 & 4.0 & -4.5 & -0.127 & -0.254 & -0.239 \\
5000 & 4.0 & -5.0 & -0.096 & -0.242 & -0.196 \\
5000 & 4.0 & -5.5 & -0.089 & -0.157 & -0.121 \\
5000 & 4.0 & -6.0 & -0.083 & -0.106 & -0.091 \\
5500 & 1.0 & -2.5 & -0.302 & -0.244 & -0.268 \\
5500 & 1.0 & -3.0 & -0.310 & -0.263 & -0.305 \\
5500 & 1.0 & -3.5 & -0.312 & -0.332 & -0.360 \\
5500 & 1.0 & -4.0 & -0.295 & -0.413 & -0.418 \\
5500 & 1.0 & -4.5 & -0.294 & -0.515 & -0.490 \\
5500 & 1.0 & -5.0 & -0.301 & -0.600 & -0.544 \\
5500 & 1.0 & -5.5 & -0.255 & -0.628 & -0.557 \\
5500 & 1.0 & -6.0 & -0.196 & -0.516 & -0.407 \\
5500 & 2.0 & -2.5 & -0.363 & -0.302 & -0.341 \\
5500 & 2.0 & -3.0 & -0.377 & -0.342 & -0.390 \\
5500 & 2.0 & -3.5 & -0.375 & -0.414 & -0.440 \\
5500 & 2.0 & -4.0 & -0.333 & -0.491 & -0.492 \\
5500 & 2.0 & -4.5 & -0.292 & -0.565 & -0.521 \\
5500 & 2.0 & -5.0 & -0.230 & -0.587 & -0.521 \\
5500 & 2.0 & -5.5 & -0.179 & -0.536 & -0.438 \\
5500 & 2.0 & -6.0 & -0.140 & -0.317 & -0.231 \\
5500 & 3.0 & -2.5 & -0.350 & -0.318 & -0.345 \\
5500 & 3.0 & -3.0 & -0.383 & -0.345 & -0.383 \\
5500 & 3.0 & -3.5 & -0.380 & -0.404 & -0.430 \\
5500 & 3.0 & -4.0 & -0.329 & -0.473 & -0.471 \\
5500 & 3.0 & -4.5 & -0.222 & -0.505 & -0.463 \\
5500 & 3.0 & -5.0 & -0.143 & -0.481 & -0.407 \\
5500 & 3.0 & -5.5 & -0.118 & -0.337 & -0.236 \\
5500 & 3.0 & -6.0 & -0.093 & -0.157 & -0.123 \\
5500 & 4.0 & -2.5 & -0.279 & -0.306 & -0.296 \\
5500 & 4.0 & -3.0 & -0.301 & -0.287 & -0.303 \\
5500 & 4.0 & -3.5 & -0.294 & -0.311 & -0.326 \\
5500 & 4.0 & -4.0 & -0.226 & -0.346 & -0.340 \\
5500 & 4.0 & -4.5 & -0.113 & -0.346 & -0.305 \\
5500 & 4.0 & -5.0 & -0.075 & -0.266 & -0.189 \\
5500 & 4.0 & -5.5 & -0.056 & -0.119 & -0.085 \\
5500 & 4.0 & -6.0 & -0.034 & -0.055 & -0.041 \\
6000 & 2.0 & -2.5 & -0.386 & -0.340 & -0.395 \\
6000 & 2.0 & -3.0 & -0.397 & -0.419 & -0.466 \\
6000 & 2.0 & -3.5 & -0.370 & -0.502 & -0.523 \\
6000 & 2.0 & -4.0 & -0.320 & -0.577 & -0.561 \\
6000 & 2.0 & -4.5 & -0.265 & -0.627 & -0.568 \\
6000 & 2.0 & -5.0 & -0.201 & -0.624 & -0.548 \\
6000 & 2.0 & -5.5 & -0.142 & -0.455 & -0.342 \\
6000 & 2.0 & -6.0 & -0.098 & -0.200 & -0.154 \\
6000 & 3.0 & -2.5 & -0.398 & -0.364 & -0.417 \\
6000 & 3.0 & -3.0 & -0.419 & -0.442 & -0.481 \\
6000 & 3.0 & -3.5 & -0.383 & -0.525 & -0.544 \\
6000 & 3.0 & -4.0 & -0.312 & -0.592 & -0.561 \\
6000 & 3.0 & -4.5 & -0.199 & -0.589 & -0.521 \\
6000 & 3.0 & -5.0 & -0.143 & -0.514 & -0.407 \\
6000 & 3.0 & -5.5 & -0.100 & -0.253 & -0.179 \\
6000 & 3.0 & -6.0 & -0.061 & -0.101 & -0.083 \\
6000 & 4.0 & -2.5 & -0.321 & -0.312 & -0.335 \\
6000 & 4.0 & -3.0 & -0.342 & -0.362 & -0.385 \\
6000 & 4.0 & -3.5 & -0.316 & -0.429 & -0.435 \\
6000 & 4.0 & -4.0 & -0.216 & -0.470 & -0.438 \\
6000 & 4.0 & -4.5 & -0.117 & -0.432 & -0.354 \\
6000 & 4.0 & -5.0 & -0.088 & -0.283 & -0.192 \\
6000 & 4.0 & -5.5 & -0.056 & -0.107 & -0.080 \\
6000 & 4.0 & -6.0 & -0.016 & -0.033 & -0.022 \\
6500 & 3.0 & -2.5 & -0.407 & -0.430 & -0.494 \\
6500 & 3.0 & -3.0 & -0.395 & -0.524 & -0.570 \\
6500 & 3.0 & -3.5 & -0.345 & -0.609 & -0.618 \\
6500 & 3.0 & -4.0 & -0.271 & -0.658 & -0.612 \\
6500 & 3.0 & -4.5 & -0.178 & -0.630 & -0.554 \\
6500 & 3.0 & -5.0 & -0.126 & -0.464 & -0.346 \\
6500 & 3.0 & -5.5 & -0.075 & -0.170 & -0.134 \\
6500 & 3.0 & -6.0 & -0.022 & -0.049 & -0.041 \\
6500 & 4.0 & -2.5 & -0.333 & -0.362 & -0.394 \\
6500 & 4.0 & -3.0 & -0.327 & -0.442 & -0.464 \\
6500 & 4.0 & -3.5 & -0.279 & -0.520 & -0.504 \\
6500 & 4.0 & -4.0 & -0.178 & -0.539 & -0.479 \\
6500 & 4.0 & -4.5 & -0.062 & -0.263 & -0.190 \\
6500 & 4.0 & -5.0 & -0.089 & -0.243 & -0.169 \\
6500 & 4.0 & -5.5 & -0.042 & -0.071 & -0.058 \\
6500 & 4.0 & -6.0 &  0.010 & -0.004 &  0.005 \\
7000 & 4.0 & -2.5 & -0.312 & -0.424 & -0.465 \\
7000 & 4.0 & -3.0 & -0.282 & -0.509 & -0.523 \\
7000 & 4.0 & -3.5 & -0.218 & -0.565 & -0.527 \\
7000 & 4.0 & -4.0 & -0.112 & -0.462 & -0.396 \\
7000 & 4.0 & -4.5 & -0.100 & -0.398 & -0.288 \\
7000 & 4.0 & -5.0 & -0.073 & -0.176 & -0.136 \\
7000 & 4.0 & -5.5 & -0.028 & -0.057 & -0.050 \\
7000 & 4.0 & -6.0 &  0.065 &  0.053 &  0.062 \\
\end{longtable}

\bibliography{papers}

\end{document}